\documentclass[conference]{IEEEtran}
\usepackage{balance}

\usepackage{amsthm}
\usepackage{amsmath}
\usepackage{mathrsfs}
\usepackage{amssymb} 
\usepackage{bbm,dsfont} 
\usepackage{multirow} 
\usepackage{units}
\usepackage{stmaryrd}   
\usepackage{verbatim}
\usepackage{enumerate} 
\usepackage[table]{xcolor}
\usepackage{slashbox} %
\usepackage{ wasysym }
\usepackage{tikz}
\usetikzlibrary{calc}
\usepackage{tabularx}

\usepackage{lipsum}
\usepackage{multicol}
\usepackage{float}

\usepackage{ifthen}

\usepackage{xcolor}
\usepackage{hyperref} %
\usepackage%
{hyperref} %
\hypersetup{
    colorlinks,
    linkcolor={blue!80!black},%
    citecolor={green!50!black},
    urlcolor={blue!80!black}
}

\usepackage{graphicx}
\graphicspath{{images/}}

\title{Bosonic Dirty Paper Coding}
\author{\IEEEauthorblockN{Uzi Pereg}\\
\IEEEauthorblockA{Institute for Communications Engineering, \\
Technical University of Munich,
 80333 Munich, Germany.\\
Email: {\tt uzi.pereg@tum.de}
 }}
\usepackage[square,numbers]{natbib}
\definecolor{light-gray}{gray}{0.8}

\definecolor{dark-gray}{gray}{0.3}
\usepackage{accents}
\newlength{\dhatheight}

\newcommand{\bieee}{\begin{IEEEeqnarray}{rCl}}
\newcommand{\eieee}{\end{IEEEeqnarray}}

\newcommand{\trace}{\mathrm{Tr}}

\newcommand{\kb}[1]{ | #1 \rangle\langle #1 | } %

\newcommand{\etal}{\emph{et al.} }

\newcommand{\etc}{etc}%

\newcommand{\ha}{\hat{a}}
\newcommand{\hb}{\hat{b}}
\newcommand{\he}{\hat{e}}

\newcommand{\hM}{\hat{M}}

\newcommand{\Aset}{\mathcal{A}}

\newcommand{\Eset}{\mathcal{E}}
\newcommand{\markovC}[1]{%
\begin{tikzpicture}[#1]%
\draw (0,0.3ex) -- (1ex,0.3ex);%
\draw (0.5ex,0.3ex) circle (0.2ex);
\draw[white] (0.2ex,0) -- (0.5ex,0);%
\end{tikzpicture}%
}
\newcommand{\Cbar}{\markovC{scale=2}}

\theoremstyle{remark}	
\theoremstyle{remark}	
\theoremstyle{remark}	
\theoremstyle{remark}	
\theoremstyle{remark} \newtheorem{definition}{Definition}
\theoremstyle{remark} 
\theoremstyle{remark}

\newcommand{\dM}{\mathsf{M}}															 	%

\newcommand{\inC}{\mathsf{C}}
\newcommand{\inR}{\mathsf{R}}

\begin{document}
\maketitle

{}

\begin{abstract} 
The bosonic channel is addressed with modulation interference and side information at the transmitter. The model can be viewed as the quantum counterpart of the classical random-parameter Gaussian channel. Based on Costa’s writing-on-dirty-paper result, the effect of the interference can be canceled. For both homodyne and heterodyne detection, we observe the same phenomenon, as the model reduces to a classical Gaussian channel. Then, we consider the bosonic channel with joint detection, for which the classical results do not apply, and derive a dirty-paper coding lower bound. We demonstrate that the optimal coefficient for dirty paper coding is not necessarily the MMSE estimator coefficient as in the classical setting.
\end{abstract}


\section{Introduction}
\begin{center}
\begin{figure}[b]
\includegraphics[scale=0.7,trim={1.5cm 12cm 8cm 12.25cm},clip]
{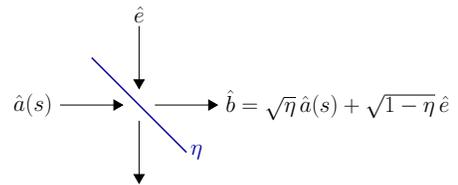} %
\caption{The beam-splitter relation of the single-mode bosonic channel.
}
\label{fig:BSp}
\end{figure}

\end{center}

Optical communication forms %
the backbone of the Internet \cite{BardhanShapiro:16p,Savov:12z,KumarDeen:14b}.
The bosonic channel is a simple quantum-mechanical model for optical communication over free space or optical fibers 
\cite{WPGCRSL:12p,WildeHaydenGuha:12p}.
 An optical communication system consists of a modulated source of photons, the optical channel, and an optical detector. %
For a single-mode bosonic channel, the channel input is an electromagnetic field mode with an annihilation operator $\ha$, and the output is another mode with the annihilation operator $\hb$. The 
input-output relation in the Heisenberg picture  is given by %
\begin{align}
\hb=\sqrt{\eta}\, \ha +\sqrt{1-\eta}\,\he
\label{eq:BosonicCr}
\end{align}
where $\he$ is associated with the environment noise and the parameter $\eta$  is %
the transmissivity, $0\leq \eta\leq 1$, which depends on the length of the optical fiber and its absorption length \cite{EisertWolf:05c} (see Figure~\ref{fig:BSp}). 
For a lossy bosonic channel, the noise mode $\he$ is in a Gibbs thermal state. %
 Modulation is performed such that the unitary displacement operator $D(\alpha)=\exp(\alpha \ha^\dagger-\alpha^* \ha)$ is applied to the vacuum state $\kb{0}$. %
The bosonic channel can be viewed as the quantum counterpart of the classical channel with additive white Gaussian noise (AWGN) \cite{WPGCRSL:12p}.

In classical communications, various %
settings of practical significance can be described by a channel $W_{Y|X}^{(s)}$ that depends on a random parameter $S\sim p_S$ when there is channel side information (CSI) available at the transmitter %
\cite{%
KeshetSteinbergMerhav:07n,%
PeregSteinberg:19p1}. %
For example,    a cognitive radio in a wireless system may be aware of the channel state and network configuration \cite{GJMS:09p}. %
Other applications include 
memory storage where the writer knows the fault locations \cite{HeegardElGamal:83p}, %
 digital watermarking \cite{MoulinOsullivan:03p}, and spread-spectrum communication  \cite{ChenWornell:01p}, 
 where the CSI represents the host data or a pseudo-random sequence to be modulated. 
The capacity of a random-parameter classical channel with CSI at the transmitter is given by \cite{HeegardElGamal:83p}
\begin{align}
C(W)=\max_{p_{U,X|S}} [I(U;Y)-I(U;S)]
\end{align}
where $U$ is an auxiliary random variable such that $U \Cbar (X,S) \Cbar Y$ form a Markov chain.  

\vspace{-0.75cm}

A  random-parameter Gaussian channel is specified by the input-output relation $Y=X+Z+S$,
with a real-valued Gaussian noise $Z\sim 
\mathcal{N}_{\mathbb{R}}(0,\sigma_Z^2)$, an additive interference $S$ known to the transmitter, and  an input power constraint $P$.
A well-known result by Costa \cite{Costa:83p} is that the capacity of the random-parameter Gaussian channel is the same as if the interference is not there, i.e. $C(W)=\frac{1}{2}\log_2\left(1+\frac{P}{\sigma_Z^2} \right)$.
Given that $S^n$ is \emph{not} known to the receiver, it is far from obvious that the interference can be canceled out without sacrificing transmission power.
 The derivation of this result is based on Costa's dirty-paper coding (DPC) strategy \cite{Costa:83p}: Set 
\begin{align}
U=X+t S
\label{eq:UtX}
\end{align} %
such that $X$ is statistically \emph{independent} of $S$. The optimal choice of the coefficient $t$ turns out to be the same as that of the %
minimum mean-square error (MMSE) estimator $\widehat{X}=tV$ for $X$ given an observation of $V=X+Z$ (see \cite[Section 4.1]{KeshetSteinbergMerhav:07n}), namely, 
\begin{align}
t=\frac{P}{P+\sigma_Z^2} \,.
\label{eq:tMMSE}
\end{align}
Explicit code constructions based on lattice codes were proposed in \cite{PhilosofErezZamir:03c,EreztenBrink:05p} and references therein.

A quantum channel with random parameters is defined by a completely-positive trace-preserving map $\mathcal{E}^{(s)}_{A\rightarrow B}$ that depends on a classical random parameter $S\sim p_S$. At first, consider finite dimensions.
Recently, the author \cite{Pereg:20c1} has determined that the capacity of the random-parameter quantum channel with CSI at the transmitter is given by the regularized formula $C(\Eset)=\lim_{n\to\infty} \frac{1}{n}\inC_1(\Eset^{\otimes n})$, with 
\begin{align}
\inC_1(\Eset)=\sup_{p_{X|S}\,,\; \sigma_A^{x,s} } [I(X;B)_\rho-I(X;S)]
\end{align}
where $I(X;B)_\rho$ is the Holevo information calculated with respect to 
$\rho_B^{x,s}= %
 \mathcal{E}^{(s)}_{A\rightarrow B}(\sigma_A^{x,s})$. 
We extend this result to the bosonic channel with infinite-dimension Hilbert spaces based on the discretization limiting argument by Guha \etal \cite{GuhaShapiroErkmen:07p}. Related settings were also considered in 
\cite{BocheCaiNotzel:16p,AnshuJainWarsi:19p,PeregDeppeBoche:21p}.

In this work, we consider the single-mode lossy bosonic channel with a coherent-state protocol and a non-ideal displacement operation in the modulation process:
\begin{align}
|\zeta_1 \zeta_2 \cdots \zeta_n \rangle =D(\alpha_1+s_1)|0\rangle %
\otimes \cdots \otimes
D(\alpha_n+s_n)|0\rangle
\label{eq:Zetan}
\end{align}
where the parameter $s_i$ represents classical interference in the transmission equipment, which the transmitter becomes aware of, while the receiver is not. %
Alternatively, this can be viewed as a watermarking model with a quantum embedding.
Given a classical host data sequence $s_1,\ldots,s_n$, %
Alice encodes an authentication message $m$ into a watermark $(\alpha_i(m,s_1,\ldots,s_n))_{i=1}^n$. Next, Alice performs a quantum embedding of the watermark; 
she prepares a \emph{watermarked state} $|\zeta_1 \zeta_2 \cdots \zeta_n \rangle$ as in (\ref{eq:Zetan}), %
and transmits it to the authenticator Bob through the optical fiber.
The capacity of the random-parameter bosonic channel represents the optimal rate at which the authenticator can recover the messages with high fidelity. %

First, we consider homodyne and heterodyne detection. Both %
reduce to a classical random-parameter channel with either real or complex-valued Gaussian noise. Thereby, we observe that based on Costa's dirty-paper solution, the effect of the classical interference can be canceled, and the capacity is the same regardless of the intensity of the interference. 
Then, we consider joint detection, in which case, the problem does not reduce to that of a classical description.
We derive a DPC lower bound based on the author's previous result \cite{Pereg:20c1}, with a general coefficient $t$ (see (\ref{eq:UtX})).
Considering the special case of a pure-loss bosonic channel, we show that the optimal coefficient is not necessarily the MMSE value as 
in (\ref{eq:tMMSE}).

\section{Definitions}
\label{subsec:notation}
 We use the following notation; %
$X$  is a classical random variable,  $p_X$ the probability density function; %
$\alpha$ is a complex-valued %
variable.  
$X\sim\mathcal{N}_{\mathbb{R}}(0,\sigma^2)$ indicates %
a real Gaussian variable, with %
$p_X(x)=\frac{1}{\sqrt{2\pi\sigma^2}}e^{-x^2/2\sigma^2}$.
  A complex Gaussian %
	$\alpha\sim\mathcal{N}_{\mathbb{C}}(0,\sigma^2)$ satisfies %
	$\alpha=X+\textrm{i}Y$ with %
	independent %
	$X,Y\sim\mathcal{N}_{\mathbb{R}}(0,\sigma^2)$. %
A quantum state of a system $A$ is denoted %
$\rho_{A}$. %
A  channel $\Eset_{A\rightarrow B}$ is a cptp %
map that maps an input $\rho_A$ to an output  $\rho_B$. A random-parameter quantum channel $\Eset^{(s)}_{A\rightarrow B}$ depends on a classical parameter $S\sim p_S$.
The channel is %
memoryless, mapping %
$\rho_{A_1 A_2\ldots A_n}$ to %
$\left(\bigotimes_{i=1}^n \Eset^{(S_i)}_{A_i\rightarrow B_i} \right)(\rho_{A_1 A_2\ldots A_n}) $, where %
$S_i$ are i.i.d. %
$\sim p_S$.
Given %
$\rho_{AB}$,
define the von Neumann entropy %
$%
H(A)_\rho = -\trace[ \rho_A\log(\rho_A) ] %
$, %
and let
$%
I(A;B)_\rho=H(A)_\rho+H(B)_\rho-H(AB)_\rho %
$. %
A detailed description of %
bosonic systems can be found in \cite{WPGCRSL:12p}. %
The vacuum state of a single-mode electromagnetic field is denoted by $%
\kb{0}$.
The \emph{creation operator} $\ha^{\dagger}$ creates an excitation: %
$\ha^{\dagger}|0\rangle=|1\rangle$, $\ha^{\dagger}|1\rangle=\sqrt{2}|2\rangle$, \etc. Whereas, %
the \emph{annihilation operator} $\ha$ takes away an excitation. %
A coherent state $|\alpha\rangle$ %
corresponds to an oscillation of the bosonic field, %
obtained by %
$|\alpha\rangle=D(\alpha)|0\rangle$, where
$D(\alpha)\equiv \exp(\alpha \ha^\dagger-\alpha^* \ha)$. %
 A thermal state with a mean photon number $N$  is given by %
$%
\tau(N)\equiv \int_{\mathbb{C}} d^2 \alpha \frac{e^{-|\alpha|^2/2N}}{\pi N} \kb{\alpha} %
$. %
A single-mode bosonic channel is described by the relation (\ref{eq:BosonicCr}) between the input and output %
operators, $\ha$ and $\hb$, %
with an input constraint %
$
N_A$. 
For a \emph{lossy} channel, the noise mode $\he$ is in a thermal state $\tau(N_E)$,  and
for a \emph{pure-loss}  channel, %
in the vacuum state. Here, we consider the lossy bosonic channel with a coherent-state protocol and  non-ideal  modulation, %
as specified in (\ref{eq:Zetan}).
The coding definitions are given below. %
\begin{definition} 
An $(\dM,n)$ code for the bosonic channel with a coherent-state protocol and CSI at the encoder consists of 
a set of $\dM$ messages,  
an encoding map  $\mathsf{Enc}:$ $ (m,s_1,\ldots,s_n) \mapsto (\alpha_i(m,s_1,\ldots,s_n))_{i=1}^n$, %
and a %
measurement $\mathsf{Dec}:$ $\rho_{B_1,\ldots,B_n} \mapsto \hM$. The codewords satisfy %
$%
|\alpha_i|^2 \leq N_A$. %
 Alice chooses a message $m$ from the message set. Using her access to the random parameters, she computes $(\alpha_i(m,s_1,\ldots,s_n))_{i=1}^n$. The modulation produces the coherent state %
in (\ref{eq:Zetan}), %
 which is then sent through the channel. %
 Bob receives the channel output systems $(B_i)_{i=1}^n$ and applies a measurement. The outcome $\hM$ is his estimate of the message $m$. %
The coding rate is defined as $R=\frac{\log_2(\dM)}{n}$ in units of bits per transmission, and the maximal probability of error is denoted by $P_{e}^{(n)}=\max_{m}\Pr(\hM\neq m|m)$.  
A rate $R>0$ is called achievable if there exists a sequence of $(\lceil 2^{nR} \rceil,n)$ codes such that %
$P_{e}^{(n)}\rightarrow 0$ %
as $n\rightarrow\infty$.
 The operational capacity $C(\Eset)$ %
is defined as the supremum of achievable rates. %
\end{definition}

\vspace{-0.05cm}
\section{Results}
First, we consider homodyne and heterodyne detection.
A homodyne measurement of a quadrature observable is implemented in practice by combining the target quantum mode with an intense local oscillator at a 50:50 beam splitter, and measuring the  photocurrent difference of the outgoing modes using two photodetectors 
\cite{BraunsteinvanLoock:05p}. 
When homodyne detection is used with a coherent-state protocol, the resulting channel $\Eset_{\text{hom}}$ is the random-parameter classical Gaussian channel
\begin{align}
Y=\sqrt{\eta}(\alpha+S)+Z_{\text{hom}}
\end{align}
with  a real-valued Gaussian %
parameter $S\sim \mathcal{N}_{\mathbb{R}}(0,N_S)$ and %
noise $Z_{\text{hom}}\sim \mathcal{N}_{\mathbb{R}}\left(0,\frac{1}{4}\left[ 2(1-\eta)N_E+1 \right] \right)$ \cite{Guha:08z}. %
Using the DPC scheme, we take $\alpha\sim \mathcal{N}_{\mathbb{R}}(0,N_A)$ and $U=\alpha+t_1 S$ with $t_1=\frac{\eta N_A}{\eta N_A+\text{var}(Z_{\text{hom}})}$, such that $\alpha$ and $S$ are uncorrelated.
The effect of the interference is thus removed, and the capacity is %
\begin{align}
C(\Eset_{\text{hom}})=\frac{1}{2}\log_2\left( 1+\frac{4\eta N_A}{2(1-\eta)N_E+1} \right) %
\end{align}
 as without interference. %

In heterodyne detection, two quadratures are measured by combining  the measured mode with a vacuum %
mode into a 50:50 beam splitter, and homodyning the quadratures of the outcome modes \cite{WPGCRSL:12p}.
Heterodyne detection is described by a random-parameter  channel $\Eset_\text{het}$ with  complex-valued Gaussian noise, specified by
\begin{align}
Y=\sqrt{\eta}(\alpha+S)+Z_{\text{het}}
\end{align}
with a complex-valued circularly-symmetric Gaussian random parameter $S\sim \mathcal{N}_{\mathbb{C}}(0,\frac{1}{2}N_S)$ and noise  $Z_{\text{het}}\sim \mathcal{N}_{\mathbb{C}}(0,\frac{1}{2}[(1-\eta)N_E+1])$ \cite{Guha:08z}. 
Similarly, we use DPC with $\alpha\sim \mathcal{N}_{\mathbb{C}}(0,\frac{1}{2} N_A)$ and $U=\alpha+t_2 S$, with $t_2=\frac{\eta N_A}{\eta N_A+\text{var}(Z_{\text{het}})}$, %
achieving the capacity
\begin{align}
C(\Eset_{\text{het}})=\log_2\left( 1+\frac{\eta N_A}{(1-\eta)N_E+1} \right) %
\end{align}
 as without interference.

For joint detection, %
the channel does not have a classical description. 
Applying the previous result by the author \cite{Pereg:20c1} for a  quantum channel with random parameters, and using the DPC strategy, %
we obtain the lower bound $C(\Eset_{\text{joint}})\geq \inR_{\text{DPC}}(t)$, %
\begin{multline}
\inR_{\text{DPC}}(t)\equiv I(\gamma;B)-I(\gamma;S) \Big|_{\gamma=\alpha+tS}
\\
=g(\eta(N_A+N_S)+(1-\eta)N_E) \qquad\qquad
\\ %
- g\left(\frac{\eta (1-t)^2 N_A N_S}{N_A+t^2 N_S}+(1-\eta)N_E \right)
\\ %
-\log_2\left( \frac{N_A+t^2 N_S}{N_A} \right)
\label{eq:DPClower}
\end{multline}
where %
$g(N)$ is the von Neumann entropy of the thermal state $\tau(N)$. That is, $g(N)=(N+1)\log_2(N+1)-N\log_2(N)$ for $N>0$, and $g(0)=0$.
The second equality in (\ref{eq:DPClower}) holds since $\zeta\equiv \alpha+S=\gamma+(1-t)S$, and the conditional variance of the channel parameter $S$ given $\gamma$ is
\begin{align*}
\text{var}(S|\gamma)=\left[1-\frac{ (\text{cov}(\gamma,S))^2 }{\text{var}(S)\text{var}(\gamma)} \right]\text{var}(S)
=\frac{N_A N_S}{N_A+t^2 N_S} \,.
\end{align*}

\vspace{-0.7cm}
\begin{center}
\begin{figure}[tb]
\includegraphics[scale=0.48,trim={-2cm 0.4cm 0 0},clip]
{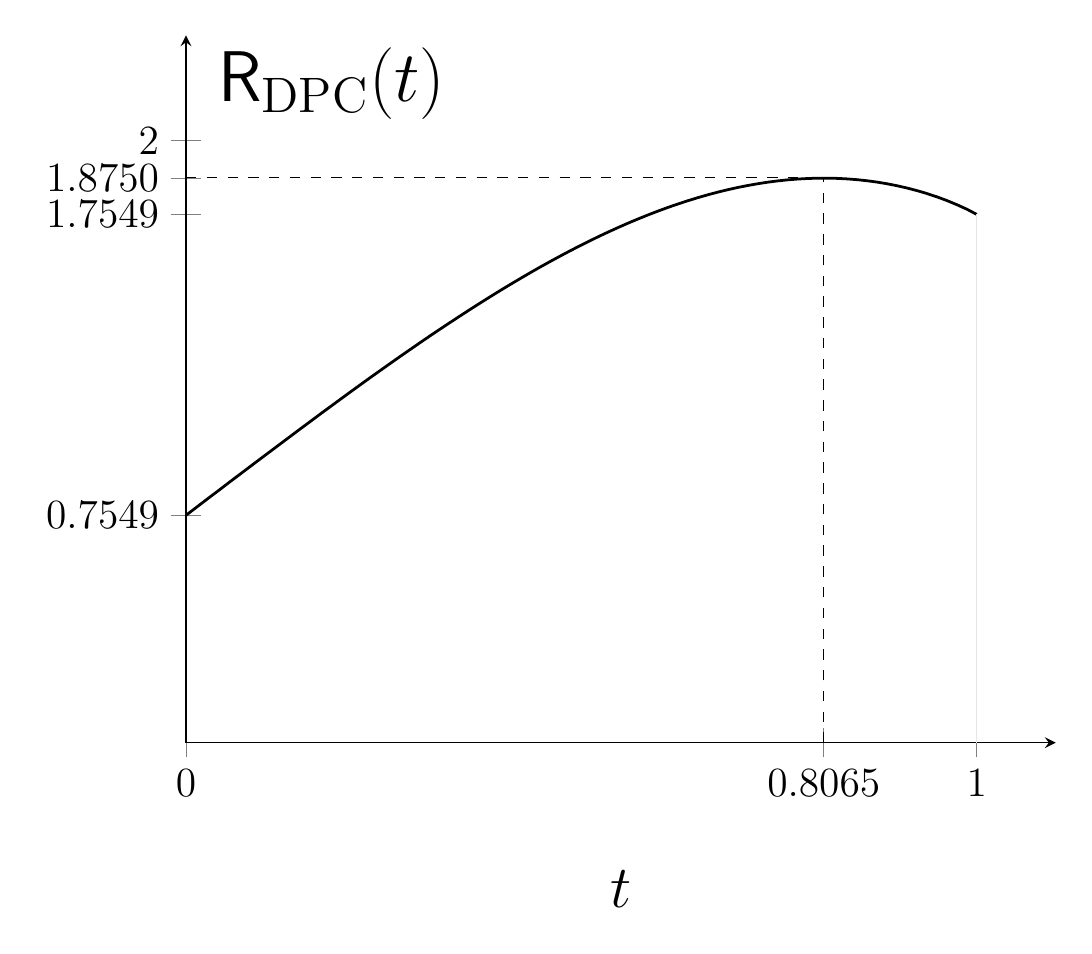} %
\caption{The DPC lower bound for the pure-loss bosonic channel with joint detection and a coherent-state protocol, for an input constraint $N_A=2$, interference mean photon number $N_S=2$, and transmissivity $\eta=\frac{1}{2}$. The bound is based on the DPC assignment of the auxiliary random variables, $\alpha\sim \mathcal{N}_{\mathbb{C}}(0,\frac{1}{2} N_A)$ and $\gamma=\alpha+tS$.
 The figure depicts the lower bound $\inR_{\text{DPC}}(t)$ as a function  of the DPC coefficient $t\in [0,1]$ (see (\ref{eq:DPCex2})).  
}
\label{fig:jDPC}
\end{figure}
\end{center}

 In particular, consider the special case of a pure-loss bosonic channel, where $N_E= 0$. 
In this case, 
\begin{align}
\inR_{\text{DPC}}(t)=&
g(\eta(N_A+N_S))- g\left(\frac{\eta (1-t)^2 N_A N_S}{N_A+t^2 N_S} \right)\nonumber\\ &-\log_2\left( \frac{N_A+t^2 N_S}{N_A} \right) \,.
\end{align}
To demonstrate, suppose that $N_A=N_S=2$ and $\eta=\frac{1}{2}$. Then, we have
\begin{align}
\inR_{\text{DPC}}(t)&=g(2)-g\left(\frac{(1-t)^2}{1+t^2}\right)-\log_2(1+t^2) \,. %
\label{eq:DPCex2}
\end{align}
Ignoring the CSI, we obtain a rate $\inR_{\text{DPC}}(t=0)=g(2)-g(1)=%
3\log_2 3-4=0.7549$. 
Whereas, using the DPC scheme with the MMSE coefficient $%
t_0=\frac{2}{2+0}=1$, we obtain a better rate:  $\inR_{\text{DPC}}(t=1)=3\log_2 3-2-\log_2 2=%
1.7549
$. The optimal value for DPC turns out to be $t_{\max}=0.8065$, for which 
\begin{align}
\inR_{\text{DPC}}(t_{\max})=1.8750 .
\end{align}
 See Figure~\ref{fig:jDPC}. The rate above is higher than the homodyne and heterodyne-detection capacities, $C(\Eset_{\text{hom}})= 1.1609$ and $C(\Eset_{\text{het}})= 1$, respectively.
However, this rate is lower than 
 the joint-detection capacity without interference ($N_S=0$), which is given by 
$g(1)=  %
2$.

Our results can be further extended to other optical channels. In particular, the random-parameter thermal amplifier channel $\Aset$ with an amplification gain $\kappa>1$ has the input-output relation
$\hb=\sqrt{\kappa}\,\ha(s)+\sqrt{\kappa-1}\,\he^\dagger$ \cite{%
GiovannettiGarciaCerfHolevo:14p}. In a similar manner, we obtain the DPC lower bound %
\begin{multline}
C(\Aset)\geq %
\max_{t\in [0,1]} \Bigg[ g(\kappa(N_A+N_S)+(\kappa-1)N_E) \qquad\qquad
\\ %
- g\left(\frac{\kappa (1-t)^2 N_A N_S}{N_A+t^2 N_S}+(\kappa-1)N_E \right)
\\ %
-\log_2\left( \frac{N_A+t^2 N_S}{N_A} \right) \Bigg].
\label{eq:DPClowerAmpl}
\end{multline}

\section{Discussion}
We conclude with the following remarks on the comparison between the classical and quantum settings:

\begin{enumerate}
\vspace{0.01cm}
\item
Costa \cite{Costa:83p} provided the intuitive analogy of `writing on dirty paper'. %
When a writer (Alice) is given a dirty paper, she knows the location and intensity of the dirt spots before writing. On the other hand, the reader (Bob) sees a mixture of the written text (channel input) and the dirt (channel parameter) without prior knowledge. %
In our setting, the dirt is the interference $s_i$ in the modulation displacement $D(\alpha_i+s_i)$. Alternatively, in the quantum watermarking scheme that we have described in the introduction, the dirt is the host covertext. 

\vspace{0.1cm}
\item
The classical capacity result can be derived using the DPC strategy in (\ref{eq:UtX})-(\ref{eq:tMMSE}) following the observation that $U-tY=X-t(X+Z)$ is the error of the MMSE estimation of $X$ given $V=X+Z$, hence it is statistically independent of the observation. Thereby,  $(U-tY)$ is jointly independent of $(V,S)$. This in turn implies that $(U-tY)$ and $Y=V+S$ are statistically independent, leading to %
 $H(U|Y)=H(U-tY)=%
H(X|V)$ which can be used in order to show that
\begin{align*} 
I(U;Y)-I(U;S)=H(U|S)-H(U|Y)=I(X;V) 
\end{align*}
(see further details in \cite{KeshetSteinbergMerhav:07n} \cite[Section 7.7]{ElGamalKim:11b}). For a bosonic channel with joint detection, we can also write the capacity in terms of $%
H(X|S)-H(X|B)_\rho$, with conditioning on the channel output. However, conditioning on a quantum system does not necessarily carry the meaning of an observation as in the classical setting \cite{HorodeckiOppenheimWinter:05p}. %

\vspace{0.1cm}
\item
While DPC was originally introduced to treat a channel with random parameters \cite{Costa:83p}, the technique is useful in multi-user setups of wireless communications as well, such as the multiple-input multiple-output (MIMO) broadcast fading channel \cite{WeingartenSteinbergShamai:06p}. It is only natural to apply and extend our results to multi-mode bosonic networks.

\end{enumerate}

\section{Acknowledgments}
This work was supported by the German BMBF %
through grant n.
16KIS0856 and by the Israel CHE Fellowship for Quantum Science and Technology.

\balance
\bibliography{references2}{}

\ifdefined\bibstar\else\newcommand{\bibstar}[1]{}\fi
\begin{thebibliography}{26}
\providecommand{\natexlab}[1]{#1}
\providecommand{\url}[1]{\texttt{#1}}
\expandafter\ifx\csname urlstyle\endcsname\relax
  \providecommand{\doi}[1]{doi: #1}\else
  \providecommand{\doi}{doi: \begingroup \urlstyle{rm}\Url}\fi

\bibitem[Bardhan and Shapiro(2016)]{BardhanShapiro:16p}
B.~R. Bardhan and J.~H. Shapiro.
\newblock Ultimate capacity of a linear time-invariant bosonic channel.
\newblock \emph{Phys. Rev. A}, 93\penalty0 (3):\penalty0 032342, 2016.

\bibitem[Savov(2012)]{Savov:12z}
I.~Savov.
\newblock \emph{Network information theory for classical-quantum channels}.
\newblock PhD thesis, McGill University, Montreal, 2012.

\bibitem[Kumar and Deen(2014)]{KumarDeen:14b}
S.~Kumar and M.~J. Deen.
\newblock \emph{Fiber optic communications: fundamentals and applications}.
\newblock John Wiley \& Sons, 2014.

\bibitem[Weedbrook et~al.(2012)Weedbrook, Pirandola, Garc\'{\i}a-Patr\'on,
  Cerf, Ralph, Shapiro, and Lloyd]{WPGCRSL:12p}
C.~Weedbrook, S.~Pirandola, R.~Garc\'{\i}a-Patr\'on, N.~J. Cerf, T.~C. Ralph,
  J.~H. Shapiro, and S.~Lloyd.
\newblock Gaussian quantum information.
\newblock \emph{Rev. Mod. Phys.}, 84:\penalty0 621--669, May 2012.

\bibitem[Wilde et~al.(2012)Wilde, Hayden, and Guha]{WildeHaydenGuha:12p}
M.~M. Wilde, P.~Hayden, and S.~Guha.
\newblock Quantum trade-off coding for bosonic communication.
\newblock \emph{Phys. Rev. A}, 86\penalty0 (6):\penalty0 062306, 2012.

\bibitem[Eisert and Wolf(2007)]{EisertWolf:05c}
J~Eisert and MM~Wolf.
\newblock Gaussian quantum channels.
\newblock In \emph{Quantum Inf. Cont. Variab. Atoms and Light}, pages 23--42.
  World Scientific, 2007.

\bibitem[Keshet et~al.(2007)Keshet, Steinberg, and
  Merhav]{KeshetSteinbergMerhav:07n}
G.~Keshet, Y.~Steinberg, and N.~Merhav.
\newblock Channel coding in the presence of side information.
\newblock \emph{Foundations and Trends in Communications and Information
  Theory}, 4\penalty0 (6):\penalty0 445--586, Jan 2007.

\bibitem[{Pereg} and {Steinberg}(2019)]{PeregSteinberg:19p1}
U.~{Pereg} and Y.~{Steinberg}.
\newblock The arbitrarily varying channel under constraints with side
  information at the encoder.
\newblock \emph{IEEE Trans. Inf. Theory}, 65\penalty0 (2):\penalty0 861--887,
  Feb 2019.

\bibitem[Goldsmith et~al.(2009)Goldsmith, Jafar, Maric, and
  Srinivasa]{GJMS:09p}
A.~Goldsmith, S.~A. Jafar, I.~Maric, and S.~Srinivasa.
\newblock Breaking spectrum gridlock with cognitive radios: An information
  theoretic perspective.
\newblock \emph{Proc. of the IEEE}, 97\penalty0 (5):\penalty0 894--914, May
  2009.

\bibitem[Heegard and Gamal(1983)]{HeegardElGamal:83p}
C.~Heegard and A.~E. Gamal.
\newblock On the capacity of computer memory with defects.
\newblock \emph{IEEE Trans. Inf. Theory}, 29\penalty0 (5):\penalty0 731--739,
  Sep 1983.

\bibitem[Moulin and O'Sullivan(2003)]{MoulinOsullivan:03p}
P.~Moulin and J.~A. O'Sullivan.
\newblock Information-theoretic analysis of information hiding.
\newblock \emph{IEEE Trans. Inf. Theory}, 49\penalty0 (3):\penalty0 563--593,
  Mar 2003.

\bibitem[Chen and Wornell(2001)]{ChenWornell:01p}
B.~Chen and G.~W. Wornell.
\newblock Quantization index modulation: A class of provably good methods for
  digital watermarking and information embedding.
\newblock \emph{IEEE Trans. Inf. Theory}, 47\penalty0 (4):\penalty0 1423--1443,
  May 2001.

\bibitem[Costa(1983)]{Costa:83p}
M.~Costa.
\newblock Writing on dirty paper (corresp.).
\newblock \emph{IEEE Trans. Inf. Theory}, 29\penalty0 (3):\penalty0 439--441,
  May 1983.

\bibitem[{Philosof} et~al.(2003){Philosof}, {Erez}, and
  {Zamir}]{PhilosofErezZamir:03c}
T.~{Philosof}, U.~{Erez}, and R.~{Zamir}.
\newblock Combined shaping and precoding for interference cancellation at low
  snr.
\newblock In \emph{Proc. IEEE Int. Symp. Inf. Theory (ISIT'2003)}, page~68,
  2003.

\bibitem[{Erez} and {ten Brink}(2005)]{EreztenBrink:05p}
U.~{Erez} and S.~{ten Brink}.
\newblock A close-to-capacity dirty paper coding scheme.
\newblock \emph{IEEE Trans. Inf. Theory}, 51\penalty0 (10):\penalty0
  3417--3432, 2005.

\bibitem[Pereg(2020)]{Pereg:20c1}
U.~Pereg.
\newblock Communication over quantum channels with parameter estimation.
\newblock In \emph{Proc. IEEE Int. Symp. Inf. Theory (ISIT'2020)}, pages
  1818--1823, Jun 2020.
\newblock URL \url{https://arxiv.org/pdf/2001.00836.pdf}.

\bibitem[Guha et~al.(2007)Guha, Shapiro, and Erkmen]{GuhaShapiroErkmen:07p}
S.~Guha, J.~H. Shapiro, and B.~I. Erkmen.
\newblock Classical capacity of bosonic broadcast communication and a minimum
  output entropy conjecture.
\newblock \emph{Phys. Rev. A}, 76\penalty0 (3):\penalty0 032303, 2007.

\bibitem[Boche et~al.(2016)Boche, Cai, and N{\"o}tzel]{BocheCaiNotzel:16p}
H.~Boche, N.~Cai, and J.~N{\"o}tzel.
\newblock The classical-quantum channel with random state parameters known to
  the sender.
\newblock \emph{J. Physics A: Math. and Theor.}, 49\penalty0 (19):\penalty0
  195302, April 2016.

\bibitem[Anshu et~al.(2019)Anshu, Jain, and Warsi]{AnshuJainWarsi:19p}
A.~Anshu, R.~Jain, and N.~A. Warsi.
\newblock On the near-optimality of one-shot classical communication over
  quantum channels.
\newblock \emph{J. Math. Phys.}, 60\penalty0 (1):\penalty0 012204, 2019.

\bibitem[Pereg et~al.(2021)Pereg, Deppe, and Boche]{PeregDeppeBoche:21p}
U.~Pereg, C.~Deppe, and H.~Boche.
\newblock Quantum channel state masking.
\newblock \emph{IEEE Trans. Inf. Theory}, 67\penalty0 (4):\penalty0 2245--2268,
  2021.

\bibitem[Braunstein and van Loock(2005)]{BraunsteinvanLoock:05p}
S.~L. Braunstein and P.~van Loock.
\newblock Quantum information with continuous variables.
\newblock \emph{Rev. Mod. Phys.}, 77:\penalty0 513--577, Jun 2005.

\bibitem[Guha(2008)]{Guha:08z}
S.~Guha.
\newblock Multiple-user quantum information theory for optical communication
  channels.
\newblock Technical report, MIT, 2008.

\bibitem[Giovannetti et~al.(2014)Giovannetti, Garcia-Patron, Cerf, and
  Holevo]{GiovannettiGarciaCerfHolevo:14p}
V.~Giovannetti, R.~Garcia-Patron, N.~J. Cerf, and A.~S. Holevo.
\newblock Ultimate classical communication rates of quantum optical channels.
\newblock \emph{Nature Photonics}, 8\penalty0 (10):\penalty0 796--800, 2014.

\bibitem[El~Gamal and Kim(2011)]{ElGamalKim:11b}
A.~El~Gamal and Y.H. Kim.
\newblock \emph{Network Information Theory}.
\newblock Cambridge University Press, 2011.

\bibitem[Horodecki et~al.(2005)Horodecki, Oppenheim, and
  Winter]{HorodeckiOppenheimWinter:05p}
M.~Horodecki, J.~Oppenheim, and A.~Winter.
\newblock Partial quantum information.
\newblock \emph{Nature}, 436\penalty0 (7051):\penalty0 673--676, 2005.

\bibitem[{Weingarten} et~al.(2006){Weingarten}, {Steinberg}, and
  {Shamai}]{WeingartenSteinbergShamai:06p}
H.~{Weingarten}, Y.~{Steinberg}, and S.~S. {Shamai}.
\newblock The capacity region of the gaussian multiple-input multiple-output
  broadcast channel.
\newblock \emph{IEEE Transactions on Information Theory}, 52\penalty0
  (9):\penalty0 3936--3964, 2006.

\end{thebibliography}

\end{document}